\documentclass[aps,twocolumn,groupedaddress]{revtex4}

\expandafter\let\csname equation*\endcsname\relax
\expandafter\let\csname endequation*\endcsname\relax
\usepackage{amsmath}

\usepackage{bm}
\usepackage{amssymb,amsfonts,latexsym,fancyhdr,graphicx,epstopdf,times,txfonts}

\usepackage{graphicx}

\usepackage{overpic}
\usepackage[english]{babel}
\usepackage{graphicx}
\usepackage {subfigure}
\usepackage[hidelinks]{hyperref}

\usepackage{xcolor}

\usepackage{verbatim}
\usepackage{lipsum}

\def\bra#1{\mathinner{\langle{#1}|}}
\def\ket#1{\mathinner{|{#1}\rangle}}

\begin{document}

\title{Irreversible Work and Orthogonality Catastrophe in the Aubry-Andr\'e model}

\author{Francesco Cosco}
\affiliation{QTF Centre of Excellence, Turku Centre for Quantum Physics,  Department of Physics and Astronomy, University of Turku, FI-20014 Turun yliopisto, Finland}

\selectlanguage{english}

\begin{abstract}
We address the statistical orthogonality catastrophe induced by a local quench in the Aubry-Andr\'e model from the perspective of nonequilibrium thermodynamics. We study the average work and the irreversible work production when quenching the impurity potential in proximity of an orthogonality event. We show how this description is able to capture the level crossings generating the orthogonality and the avoided crossings which causes the plateau-like structures, signature of the Aubry-Andr\'e spectrum, when considering the full statistics of orthogonality events.
\end{abstract}

\maketitle

\section {Introduction}

Introducing a small perturbation in a many-body fermionic system can give rise to a rich plethora of phenomena.
One of the more striking effect is the phenomenon  referred to as Anderson Orthogonality Catastrophe (AOC) \cite {anderson1967}.
In a nutshell, the phenomenon consists in predicting a power law decay, in the system size, for the overlap between the 
many-body ground states of a system of non-interacting fermions with ($|\Psi_0(\epsilon) \rangle$) and without ($|\Psi_0 \rangle$)  an impurity potential, namely
\begin{equation}
F \equiv |\langle \Psi_0 |\Psi_0(\epsilon) \rangle |\sim L^{-\alpha},
\label{oc1}
\end{equation}
where $\epsilon$ denotes the perturbation strength and $L$ is the size of the system. % and $\gamma$ a constant that depends on $\epsilon$. what he called infrared catastrophe for a Fermi gas perturbed by a local scattering potential . 
The consequences of AOC are  witnessed in different areas of physics such as the singular behaviour of the energy  excitation  spectra, revealed, e.g., by x-ray photoemission spectroscopy \cite {tanabe1985} or the Kondo effect in graphene \cite{hentschel2007}.  In the past few years, the phenomenon has found a new and fertile ground of studies in the controllable domain of ultracold trapped gases \cite {goold2011,knap2012,sindona2013,campbell2014}. 
More recently, a new idea of a statistical orthogonality catastrophe (StOC) has been introduced for insulating systems \cite {khemani2015,deng2015,coscostoc}. Specifically, this form of orthogonality catastrophe was highlighted studying the effect of local perturbations on the Anderson insulator and on the localised phase of the Aubry-Andr\'e model, both sharing exponentially localised single-particle eigenstates. In these two cases it has been found that the typical wave function overlap  decays exponentially with the system size
\begin{equation}
F_\textrm{typ}  \equiv \exp(\overline {\log F}) \sim  \exp (-\gamma L),
\label{stoc}
\end{equation}
where  the bar denotes an average over different realisations of the Fermi system: in the case of the Anderson insulator the average is taken over random realisations of the disorder while in the Aubry-Andr\'e the average is taken over random realisation of the phase factor. In \cite {deng2015} was also shown how the phenomenon persists also in presence of interaction between the fermions in the lattice.

\begin{figure*}[!t]
\begin {center}
\begin{overpic}[scale=0.7] {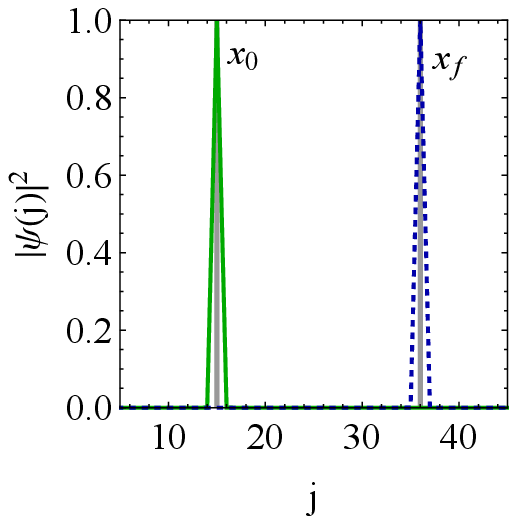}
\put(55,102){(a)}
\end{overpic}
\hspace{4 mm}
\begin{overpic}[scale=0.7] {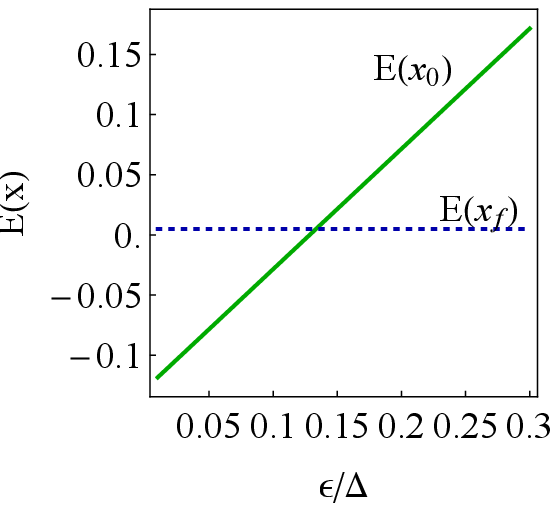}
\put(55,94){(b)}
\end{overpic}
\hspace{4 mm}
\begin{overpic}[scale=0.7] {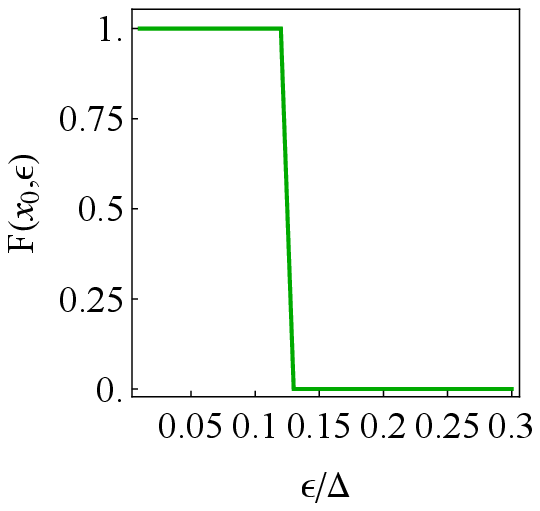}
\put(55,98){(c)}
\end{overpic}
%\begin{overpic}[scale=0.7] {./pics/psi01}
%\end{overpic}

%\hspace{2 mm}
%\begin{overpic}[scale=0.7] {./pics/f012}
%\end{overpic}
\vspace{4 mm}

\begin{overpic}[scale=0.7] {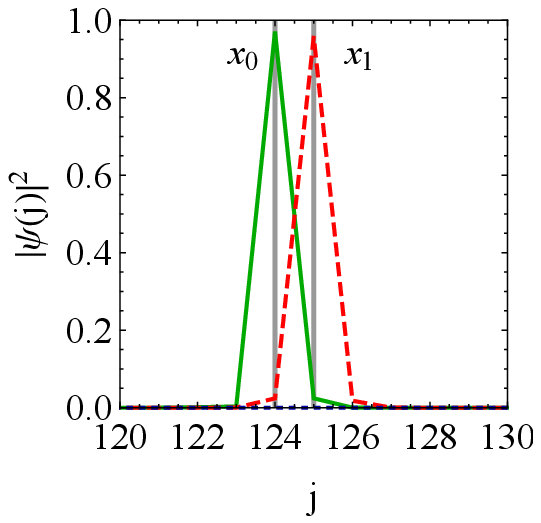}
\put(55,102){(d)}
\end{overpic}
\hspace{2 mm}
\begin{overpic}[scale=0.7] {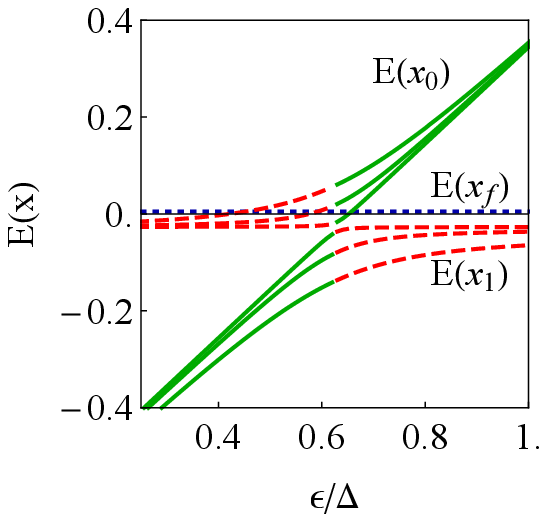}
\put(55,98){(e)}
\end{overpic}
\hspace{2 mm}
\begin{overpic}[scale=0.7] {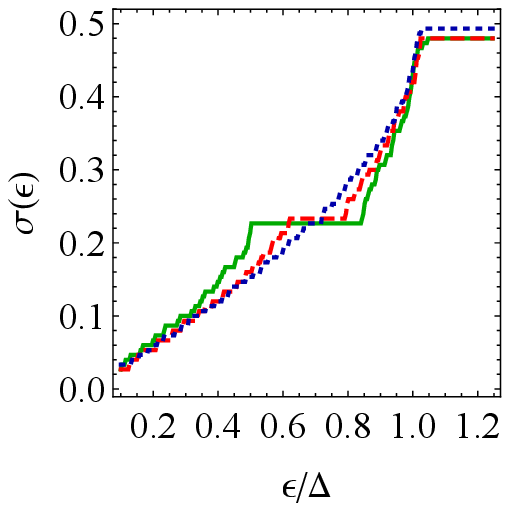}
\put(55,106){(f)}
\end{overpic}
\hspace{2 mm}
\begin{overpic}[scale=0.7] {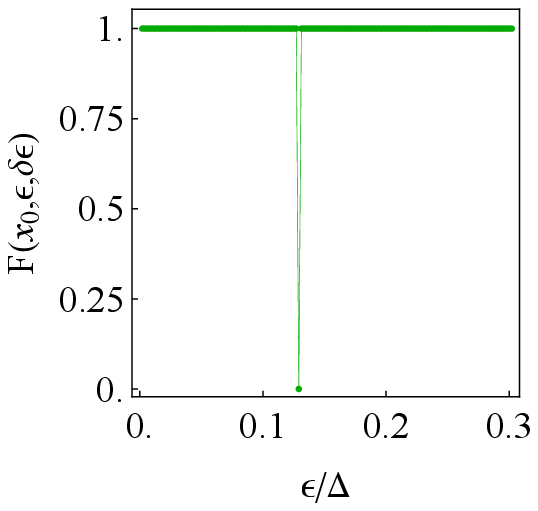}
\put(55,102){(g)}
\end{overpic}
\end{center}
\caption{Panel (a): single-particle eigenstate with no occupied neighbours localised at site $x_0$ and lowest energy unoccupied state localised at site $x_f$. Panels (b) and (c):  level crossing and fidelity when perturbing the site $x_0$ as a function of the perturbation potential. Panel (d): occupied nearest neighbours eigenstates localised at sites $x_0$ and $x_1$. Panel (e): avoided crossing  when perturbing a site with an occupied neighbour;  in dashed red and solid green the energy of the states localised on the different sites; the different pairs of curves correspond to $J/\Delta=0.1$, $0.05$ and $0.01$, which determines the gap of the avoided crossing. Panel (f): probability $\sigma$ that the ground states with and without impurity become orthogonal; $\sigma$, calculated with $\delta=10^{-2}$, is displayed versus $\epsilon/\Delta$ for $J/\Delta = 0.01$, $0.05$, $0.1$ in dotted blue, dashed red and solid green respectively. Panel (g):  fidelity of the two ground states with impurity placed at site $x_0$ of panel (a) with strength $\epsilon$ and $\epsilon+\delta \epsilon$, where $\delta \epsilon \simeq 2 \cdot 10^{-3} \Delta$. In all the panels the lattice consists of 150 sites.}
\label{thermo-stoc-plot}
\end {figure*}

\section {Orthogonality Catastrophe in Aubry-Andr\'e}
In this work we consider exclusively  a one-dimensional gas of spinless  fermions trapped  in a quasi-periodic optical lattice, known as Aubry-Andr\'e model \cite{aubry1980,sokoloff1981,thouless1983,jitomirskaya1999,modugno2009}, and described by the following Hamiltonian
\begin{equation}
\hat{H}_{AA}= -J \sum_{i=1}^{L-1} (\hat a^\dagger_{i+i} \hat a_i+\hat a^\dagger_{i} \hat a_{i+1}) +\Delta \sum_{i=1}^L   \hat n_i \cos (2 \pi \beta i + \phi), 
\label{aaham}
\end{equation}
where $J$ is the hopping parameter, $\Delta$ the the strength of the on-site potential, $\beta$ is the ratio between the frequencies of the two optical potentials generating the lattice,  $\phi$ is an arbitrary phase, and $L$ is the number of lattice sites. The hopping parameter and the on-site potential can be derived from the local forces and potentials acting on the atoms \cite {settino2017} and it is the interplay between these two that determines the phase of the system. In fact, when $\beta$ is irrational,  for  $\Delta > 2 J$ the model shows a transition from delocalized to localized single particle eigenstates. Therefore, when compared with usual Anderson localisation, in this setup  localisation emerges in a fully controllable and tunable way as successfully proven experimentally  \cite {sanchez-palencia2010,schreiber2015,luschen2017}. We take $\beta = \frac {1+\sqrt{5}}{2}$ and consider exclusively the localised phase of the model. The system is perturbed by introducing a local impurity into the Hamiltonian
through an on-site potential. As a consequence, the total Hamiltonian becomes
\begin {equation}
\hat H (x,\epsilon) =\hat H_{AA} + \epsilon \hat a _x^\dagger \hat a_x,
\label {ham-tot}
\end {equation}
where $\epsilon$ is the strength of the impurity potential and $x$ is the index labelling the site to which the perturbation potential is added. We define the  ground states overlap by taking  into explicit consideration the lattice site $x$ to which the impurity is coupled and the strength of the perturbation  as
\begin {equation}
F(x,\epsilon) \equiv |\langle \Psi_0 |\Psi_0(x,\epsilon) \rangle |,
\label {fidelity-stoc}
\end{equation}
where $\ket {\Psi_0}$ and $\ket {\Psi_0(x,\epsilon}$ are the ground states of the Hamiltonians in Eq. \eqref{aaham} and Eq. \eqref{ham-tot} respectively. Throughout this work every ground state considered is calculated at half filling, meaning that the number of fermions in the system is half the number of lattice sites.

What it is found is that for certain realisations $F \simeq 1$, while for certain others $F \simeq 0$, already at finite size. This peculiarity is the reason why this form of orthogonality catastrophe has been dubbed as $statistical$. Interestingly, the statistics of orthogonality events, when explored as a function of the perturbation potential, is able to highlight the unusual features of the Aubry-Andr\'e spectrum. For this purpose it is convenient to introduce a function able to quantify the amount of orthogonality events, i. e. 
\begin{equation}
\sigma (\epsilon) =  \frac{1}{L} \sum_{x=1}^{L} \theta(\delta - F(x,\epsilon)) ,
\label{orto-measure}
\end{equation}
which, to some extent, can be interpreted as the probability of obtaining an orthogonality event when averaging over  the position of the impurity potential. $\delta$ in Eq. \eqref {orto-measure} is a tolerance value within which we define two many-body wavefunctions to be $orthogonal$. From an operative point of view this tolerance value is needed to distinguish between the cases in which $F \simeq 1$ and $F \simeq 0$. 

Previous works have outlined how in the presence of an orthogonality event the system witnesses a change in the ground state equilibrium density, with a particle occupying a different site in the lattice in the perturbed and unperturbed configurations \cite {khemani2015,deng2015,coscostoc}. Being the system localised the arrangement is due to the fact that the impurity, altering the energy of the perturbed site, makes a different state, localised on a different site, energetically favourable.

\begin{figure*}[!t]
\begin {center}
\begin{overpic}[scale=0.7] {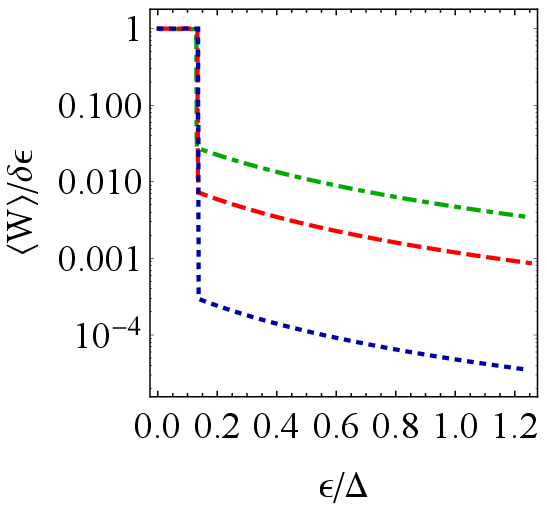}
\put(55,98){(a)}
\end{overpic}
\hspace{2 mm}
\begin{overpic}[scale=0.7] {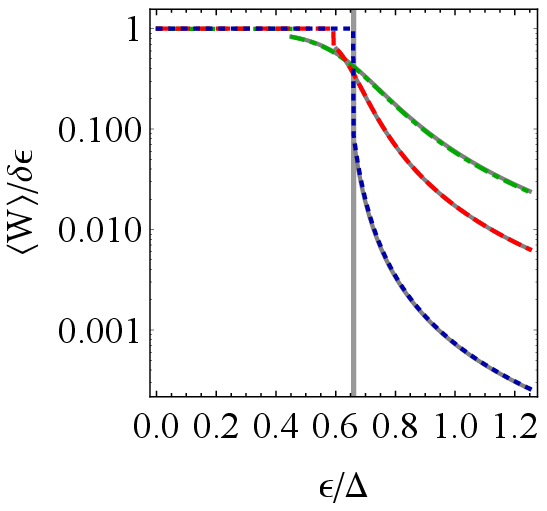}
\put(55,98){(b)}
\end{overpic}
\hspace{2 mm}
\begin{overpic}[scale=0.7] {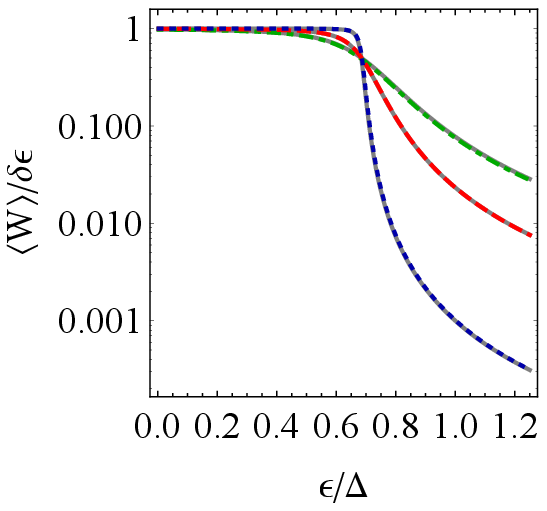}
\put(55,98){(c)}
\end{overpic}

\vspace{6 mm}

\begin{overpic}[scale=0.7] {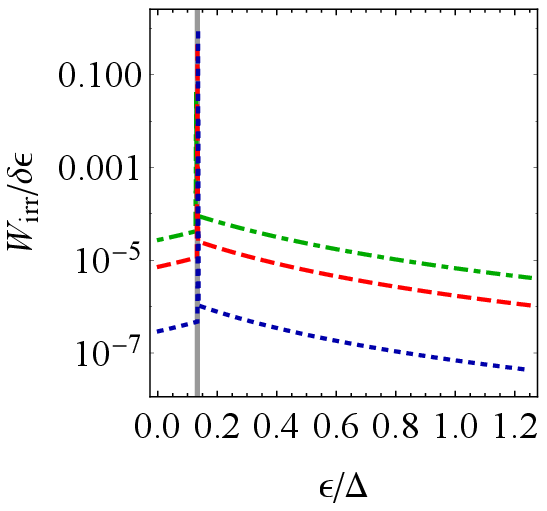}
\put(55,98){(d)}
\end{overpic}
\hspace{2 mm}
\begin{overpic}[scale=0.7] {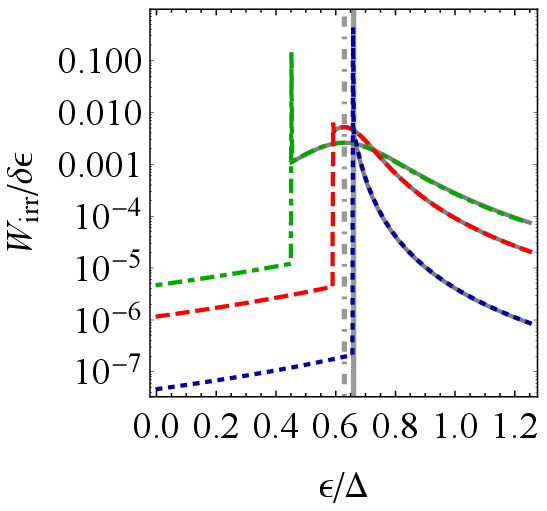}
\put(55,98){(e)}
\end{overpic}
\hspace{2 mm}
\begin{overpic}[scale=0.7] {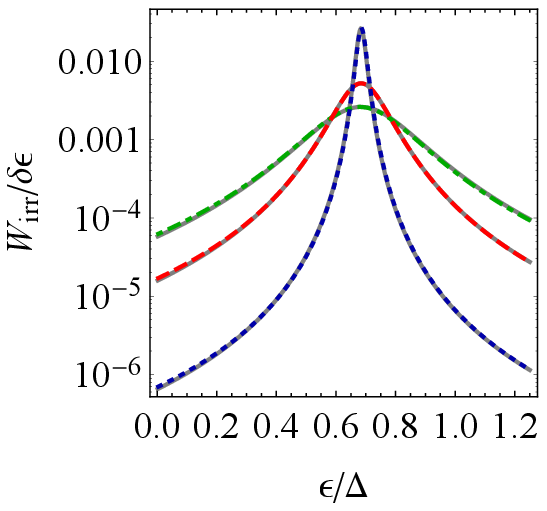}
\put(55,98){(f)}
\end{overpic}
\end{center}
\caption{Average work done (a-b-c) and Irreversible work (d-e-f) produced during an infinitesimal quench of the impurity potential. Panel (a-d), (b-e) and (c-f) correspond to the cases in which the impurity is placed in an occupied site with  no occupied neighbours, in an occupied site with an occupied neighbour, and the occupied site neighbour to $x_f$ respectively. $J/\Delta=0.1$, $0.05$ and $0.01$ are displayed in dotdashed green, dashed red and dotted blue respectively. The gray curves in (b-c) and (e-f) are obtained plotting the average work defined in Eq. \eqref {awork-a-crossing} and the irreversible work defined in Eq. \eqref {work-a-crossing}. The dashed gray vertical line in (e) and the solid gray vertical line in (b-e) correspond to $\epsilon= V_0-V_1$ and $\epsilon= V_f-V_0$ respectively.}
\label{iwork-stoc}
\end {figure*}

Within this framework we can distinguish two cases. In the first case the impurity perturbs
an occupied site, labelled as $x_0$, with no nearest neighbours occupied. Increasing the strength of the perturbation lead to an increase to the energy associated to the state localised to this site, defined as $E(x_0)$.
Therefore, an orthogonality event occurs whenever a  state, localised at $x_f$ and with energy $E(x_f)$, becomes energetically favourable to be occupied. In other words, the orthogonality is achieved when the two states, as effect of the perturbation, undergo a level crossing, i.e. $E(x_0)>E(x_f)$.
This level crossing implies a non-local ground state density rearrangement given the fact that,  as seen in panel (a) of Fig. \ref {thermo-stoc-plot}, the two corresponding eigenstates are, in general, spatially separated. Then, as a consequence of the orthogonality event, a fermion can be adiabatically moved to a different site in the lattice by means of a local perturbation. On average, this distance is found to scale linearly with the system size and not to be affected by the localisation length \cite {deng2015,coscostoc}. Panels (b) and (c) of Fig. \ref {thermo-stoc-plot} summarises this explanation of the orthogonality catastrophe showing  how  $E(x_0)$ increases when increasing the perturbation potential and  the ground state fidelity undergoes to an abrupt change in proximity of the aforementioned level crossing.  When the perturbed occupied site has an occupied neighbour, as in panel (d) of Fig. \ref {thermo-stoc-plot}, let's call it
 $x_1$, with $E(x_1)$ its related energy, the phenomenology is slightly modified and the mechanism that generates an orthogonality event changes. Although we are considering the localised phase of the Aubry-Andr\'e potential  the impurity is able to modify both $E(x_0)$ and $E(x_1)$ giving rise to an avoided crossing because of the non-zero tunnelling. Whenever one of the two is now greater than $E(x_f)$ an orthogonality event is obtained. Panel (c) of Fig. \ref {thermo-stoc-plot} shows as  for bigger $J$ the level crossing generating the orthogonality is achieved far smaller values of the perturbation. It is worth to point out that for the sake of discussion we have labelled the occupied sites as $x_0$ and $x_1$ but, especially close to the avoided crossing or when the difference between the values of the on-site potential is smaller than the tunnelling $J$, the two states are actually delocalised on the two sites. It goes without saying that in this case we are also considering the perturbed site to have a lower energy than its neighbour.

The full  statistics of orthogonality events, averaging  over the perturbation position, can be studied trough the $\sigma(\epsilon )$ function defined in Eq. \eqref{orto-measure}. Panel (f) shows how the  function accounting for the amount of orthogonality events increases monotonically for increasing values of the perturbation. However, we can notice  the emergence of a plateau-like structure with amplitude of order $2J$. This feature is originated by the avoided crossing mechanism
 which induces a level crossing with the free state at lower values of the perturbation when increasing the kinetic term. The plateau-like structure is  related to the specific properties of the Aubry-Andr\'e spectrum and its gap, which guarantee that states on opposite sides of the principal gap of order $2J$ are nearest neighbours \cite {thouless1983,coscostoc}, or rather delocalised on neighbouring sites.

\section {Average work and irreversible work production}
The picture describing the statistical orthogonality catastrophe in terms of level crossings  naturally suggests to approach the phenomenon using the tools from quantum thermodynamics and treating it, effectively, as a quantum phase transition. Recently, tools from quantum information, such ground state fidelity or fidelity susceptibility, and quantities from non-equilibrium thermodynamics such as average work and irreversible work have been used in synergy in order to explore quantum phase transitions  \cite {dorner2012,mascarenhas2014,sharma2015,bayocboc2015,bayat2016,paganelli2016,coscocoulomb,rossini2018} and Fermi gases quenched by local impurities \cite {sindona2014}. The role of these thermodynamic quantities in the studies of quantum phase transitions at zero temperature is easily understood once the connection with the the derivatives of the ground state energy is established \cite {mascarenhas2014, paganelli2016}.  In all these cases the quench under investigation is an infinitesimal change of the parameter driving the phase transition. Therefore, it is crucial to see what happens when the impurity strength is quenched by an infinitesimal amount. For this purpose we define the fidelity
\begin {equation}
F(x,\epsilon, \delta  \epsilon) \equiv |\langle \Psi_0 (x,\epsilon) |\Psi_0(x, \epsilon +\delta \epsilon) \rangle |,
\label {fidelity-work}
\end {equation}
which represents the overlap between the ground state of the two Fermi systems with strength of the impurity potential differing of an infinitesimal amount $\delta \epsilon$. The two ground states overlap of Eq. \eqref{fidelity-stoc} and Eq. \eqref{fidelity-work} behave in a dramatically different way. Indeed, when increasing the impurity potential strength $\epsilon$ the overlap of Eq. \eqref{fidelity-stoc}, as displayed in panel (a) of Fig. \ref {thermo-stoc-plot}, shows a sharp transition at the level crossing. Instead, the overlap of Eq. \eqref{fidelity-work}, as displayed in panel (g) of Fig. \ref {thermo-stoc-plot}, results in $F\simeq 0$ only  in correspondence of the level crossing. Our goal is to understand then what information we can get from a thermodynamic approach, based on quantifying the average work and the irreversible work, when such a quench is performed in proximity of this "critical" point.

The average work done on a quantum system during a sudden quench is given by the average on the initial state of the difference between  final and initial Hamiltonian \cite {talkner2007}. For an Hamiltonian written as $\hat H (\epsilon) =\hat H_0 + \epsilon \hat V$ the average work done associated to a quench in $\epsilon$, from an initial value $\epsilon_i$ to a final one $\epsilon_f$, at zero temperature is 
\begin {equation}
\langle W \rangle =   \langle \hat H (\epsilon_f)-\hat H (\epsilon_i) \rangle \simeq \delta \epsilon \frac { \partial E_{GS}(\epsilon)}{\partial \epsilon}\biggr \rvert_{\epsilon_i},
\label {work-der}
\end{equation}
where $\delta \epsilon = \epsilon_f -\epsilon_i$.  The last equality which connect the average work to the derivative of the ground state energy $E_{GS}$ is obtained  by considering $\delta \epsilon$ to be small, allowing us to use $\hat V =\partial \hat H(\epsilon)/\partial \epsilon $ and the Hellmann-Feynman theorem \cite {mascarenhas2014}.  The irreversible work produced during the quench is obtained by subtracting the free energy difference  to the average work \cite {plastina2014}, giving
\begin {equation}
W_{irr}= \delta \epsilon \frac { \partial E_{GS}(\epsilon)}{\partial \epsilon}\biggr \rvert_{\epsilon_i}-E_{GS}(\epsilon_i)+E_{GS}(\epsilon_f) \simeq - \frac { \delta \epsilon^2}{2} \frac { \partial^2 E_{GS}(\epsilon)}{\partial \epsilon^2}\biggr \rvert_{\epsilon_i},
\label {wirr-der}
\end{equation}
\begin{figure*}[!t]
\begin {center}
\begin{overpic}[scale=0.7] {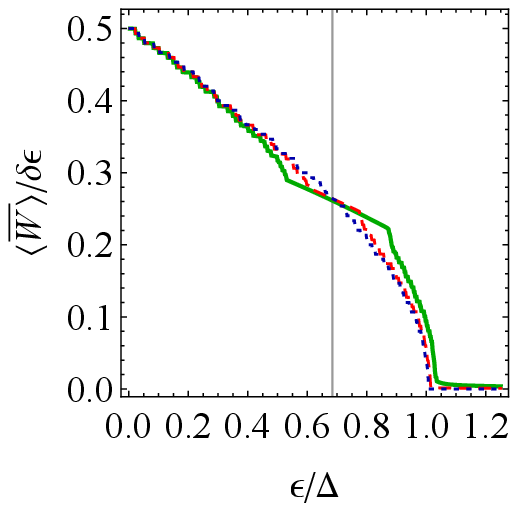}
\put(55,104){(a)}
\end{overpic}
\hspace {2mm}
\begin{overpic}[scale=0.7] {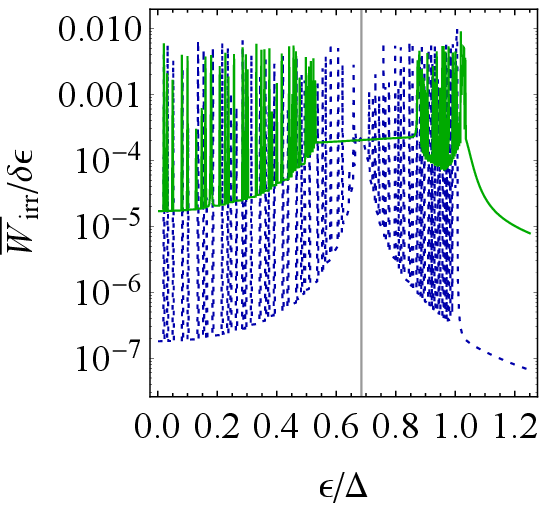}
\put(55,98){(b)}
\end{overpic}
\hspace {2mm}
\begin{overpic}[scale=0.7] {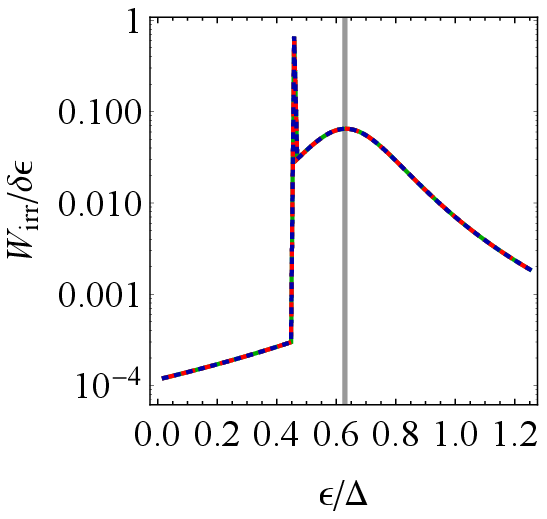}
\put(55,98){(c)}
\end{overpic}

\end{center}
\caption{Panel (a): average work  averaged over the impurity position, and in solid green, dashed red, and dotted blue are displayed the results for $J/\Delta=0.1$, $0.05$ and $0.01$, respectively. Panel (b): the irreversible work  averaged over the impurity position, and in solid green and dotted blue are displayed the results for $J/\Delta=0.1$ and $0.01$, respectively. Panel (c): irreversible work produced  for the case in which the impurity is placed in a site with an occupied neighbours for different lattice sizes, namely $N_s=150$, $300$, and $450$ in green, red, and blue, respectively. The vertical lines in (a) and (b)  are at $\epsilon = V_f-V_0$. The vertical line in (c) is at $\epsilon = V_1-V_0$.}
\label{iwork-avg-stoc}
\end {figure*}
where again the last equality holds in the case of sudden infinitesimal quenches.
In the case we are studying is trivial to see how the average work done on the system when quenching the impurity potential on a fixed site is proportional to the equilibrium occupation of the quenched site, i. e.
\begin {equation}
\langle W \rangle = \delta \epsilon   n (x, \epsilon),
\end {equation}
where  $n (x,\epsilon)= \langle \Psi_0 (x,\epsilon)| \hat n_x |\Psi_0(x, \epsilon) \rangle$.

\section {Orthogonality catastrophe and work statistics}
In this section we present the results obtained by calculating the average work done and the irreversible work production produced during sudden and infinitesimal quenches of the impurity potential added to the Aubry-Andr\'e Hamiltonian. We focus our attention on three possible positions for the impurity.  They represent all the possible cases obtainable when perturbing a system of fermions trapped in a quasi-periodic potential at half filling with a local quench. In the first case, the impurity is placed  in an occupied site with no nearest neighbours occupied. In the second case, the impurity is place  in an occupied site with an occupied nearest neighbour. The third case corresponds to the unique case in which the impurity is placed in the occupied site nearest neighbour of the lowest energy unoccupied site of the Hamiltonian without the impurity.

Panels (a), (b) and (c) of Fig. \ref{iwork-stoc} display the average work done in the aforementioned three cases respectively. The average work done captures in all the three cases the change in the occupation of the quenched site at the level crossing. Reducing the hopping factor $J$ the transition gets sharper and sharper in the cases of quenched site with occupied neighbours and quenched site nearest neighbour to the lowest energy unoccupied site. In the limit $J \rightarrow 0$, after the level crossing, the occupation of the perturbed site and the average work vanish. 

We proceed displaying in panels (d), (e) and (f) of  Fig. \ref{iwork-stoc}  the irreversible work produced in the same three cases. Interestingly,  the irreversible work  is able to capture both  the level crossing and the avoided crossing. In detail, in panel (d) of Fig \ref{iwork-stoc} is shown the irreversible work when perturbing a site with unoccupied neighbours. In this case the irreversible work is discontinuous, as expected at the level  crossing, and it is unaffected by different tunnellings, as soon as we remain in the strongly localised phase. Panel (e), corresponding to the quench of a site with an occupied neighbour, displays more interesting features. The irreversible work is  discontinuous at the level crossing, between $E(x_1)$ and $E(x_f)$, and at the avoided crossing, between  $E(x_0)$ and $E(x_1)$, it displays a local maximum. By reducing the tunnelling $J$ the discontinuity moves to bigger values of the perturbation, in line with the shift displayed in panel (f) of Fig. \ref {thermo-stoc-plot}. It is worth to recall that when we quench a site with an occupied neighbour we consider to place the impurity in the site with the smaller value of the on-site potential, corresponding then to the state with a lower energy in the single-particle spectrum. The role of the avoided level crossing can be further corroborated by introducing an  effective two level system to describe these many-body thermodynamic quantities. Let us define the two level Hamiltonian
\begin {equation}
\hat H_2(\epsilon) =  
\begin{bmatrix}
    V_0 + \epsilon       & -J \\
    -J       & V_1 ,
\end{bmatrix} 
\label {h2-model}
\end {equation}
where $V_0 = \Delta \cos (2 \pi \beta x_0+\phi)$ and $V_1 = \Delta \cos (2 \pi \beta x_1+\phi)$ are the values of the on-site potential on the quenched site $x_0$ and its nearest neighbour $x_1$. By calculating the ground state energy of the simple model in Eq. \eqref {h2-model},  and using the expression given in Eq. \eqref {work-der}, we find that the average work done is
\begin{equation}
\langle \tilde W \rangle =\delta \epsilon \frac{{V_1}-{V_0}-\epsilon+\sqrt{4 J^2+({V_0}-{V_1}+x)^2}}{2 \sqrt{4 J^2+({V_0}-{V_1}+\epsilon)^2}}.
\label {awork-a-crossing}
\end {equation}
With a second derivative in the impurity potential strength of the ground state energy of Hamiltonian in Eq. \eqref {h2-model},  as in Eq. \eqref {wirr-der}, we find the irreversible work produced during the sudden quench  to be
\begin{equation}
\tilde W_{irr} = -\frac{\delta \epsilon^2 J^2}{\left(4 J^2+(V_0-V_1+\epsilon)^2\right)^{3/2}}.
\label {work-a-crossing}
\end{equation} 
The two-level description provides an optimal fit after the discontinuity as depicted by the grey curves in panels (b) and (e) of Fig. \ref {iwork-stoc}. A particular and unique case is when the perturbed occupied site is the neighbour of $x_f$, the lowest energy unoccupied state of the Hamiltonian without the impurity. In this case the irreversible work does not present any discontinuity and at non-zero tunnelling the average work and the irreversible work are completely and perfectly described by the expressions in Eq. \eqref {awork-a-crossing} and  Eq. \eqref {work-a-crossing}, with $x_1=x_f$, as displayed with the grey solid curves in panels (c) and (f) of Fig. \ref{iwork-stoc}.

Finally, to better compare the thermodynamic approach to the full statistics of orthogonality events   we consider now to perform the average over the position of the impurity in the lattice. The average work done and the irreversible work become $\langle \overline W \rangle = 1/L \sum_{x=1}^L \bra {\Psi (x, \epsilon)}\hat H(x,\epsilon+\delta \epsilon)-\hat H(x,\epsilon)\ket {\Psi (x, \epsilon)} $ and consequently $\overline W_{irr}=\langle \overline W \rangle-1/L \sum_{x=1}^L (E_{GS} (x,\epsilon)-E_{GS} (x,\epsilon+\delta \epsilon))  $, where $E_{GS} (x,\epsilon)$ is the many-body ground state energy of Hamiltonian in Eq. \eqref {ham-tot} at half filling. This mean average work for $\epsilon = 0 $ coincides with the filling factor when divided by the infinitesimal amount $\delta \epsilon$, and in our case we obtain $\langle \overline W \rangle/ \delta \epsilon =1/L \sum_{x=1}^L n(x,0)=1/2$. By increasing $\epsilon$ the average work decreases and in correspondence of the plateau of the $\sigma$ function, as in panel  (f) of Fig. \ref{iwork-stoc},
 we witness the emergence of a region in which the work decreases with a different slope as depicted in panel (a) of Fig. \ref {iwork-avg-stoc}. More interesting is the behaviour of the mean irreversible work. In fact, it reveals all the discontinuities related to all the possible level crossings induced in the single particle spectrum, with the developing of a region with no peaks, as shown in panel (b) of Fig. \ref{iwork-avg-stoc}. This region of no peaks is the analogous of the plateau of panel (f) of Fig. \ref {thermo-stoc-plot} and beautifully captures the fact that in this range of interaction no level crossing is induced.  By reducing the tunnelling the region shrinks and its centre coincides with the difference of the on-site potential between the lowest energy unoccupied site and its occupied neighbour, i.e., at half filling, it is found $\epsilon = V_f-V_0 \approx \Delta |\sin (2 \pi \beta)|$.

To conclude this section, it is worth to discuss the role of the lattice size in these thermodynamics quantities. We have found, as expected for a localised system with a gapped spectrum, no noticeable changes  for the single site realisation. In fact, panel (c) of Fig. \ref{iwork-avg-stoc} shows a perfect overlap for the irreversible work produced when quenching the same single site with an occupied neighbour for increasing values of the lattice length in the range of sizes considered. Further increasing the size will eventually result in a different energy for the lowest energy unoccupied state, i.e $E(x_f)$. On average, and for bigger sizes, at half filling $E(x_f) \sim \frac {1}{L}$. Furthermore, when the lattice size is increased an increasing number of discontinuities can be witnessed in the  irreversible work when averaging over the impurity position. This behaviour signals the fact that  increasing the number of energy levels in the occupied spectrum consequently, and unsurprisingly, increases the number of possible  level crossings.

\section {Conclusions}

In this work we have employed an approach based on thermodynamic quantities, such as average work and irreversible work production, to characterize the novel phenomenon of statistical orthogonality catastrophe in the localised phase of the non-interacting fermionic Aubry-Andr\'e model. To this purpose, we have considered sudden and infinitesimal quenches of the impurity potential. In this framework the ground state overlap vanishes only at the level crossing between the modified energy of the perturbed site and the energy of the lowest energy unoccupied state of the unperturbed Hamiltonian. Both average work and irreversible work are able to signal the level crossing with an abrupt change, but the irreversible work is also able to witness the avoided crossings induced when quenching an occupied site with an occupied neighbour. Furthermore, the description of the avoided crossing mechanism can be further corroborated  by the description in terms of an effective two-level system model.

 To conclude, we have shown how a non-equilibrium thermodynamic approach, usually employed to the study of quantum phase transitions, can be applied to study the statistical orthogonality catastrophe and it is able to reveal new insight on the rich physics obtained when quenching with a local impurity an Aubry-Andr\'e Hamiltonian. Beyond the fundamental interest, this critical behaviour around the orthogonality catastrophe "critical" points might be harnessed to design an impurity driven quantum Otto engine, for which it has been proven how a working substance close to criticality can lead to improved and enhanced performances \cite {campisi2016}.

\section* {ACKNOWLEDGEMENTS}
F.C. acknowledges support from the Horizon 2020 EU col- laborative projects QuProCS (Grant Agreement No. 641277) and the Academy of Finland Centre of Excellence program (Project no. 312058).
  \bibliographystyle{apsrev}%{unsrt}%{utphys}%{/home/sarah/apj} % % {elsart-num} % {apsrev} % {mn2e} %{apj}
  % looks for "bibliography.bib" in $BIBINPUTS, so make sure your $BIBINPUTS is correctly set. This example
  % WILL NOT work without bibliography.bib!
 % \bibliography{text/bibliography}
  \bibliography{thesisbib}

\begin{thebibliography}{31}
\expandafter\ifx\csname natexlab\endcsname\relax\def\natexlab#1{#1}\fi
\expandafter\ifx\csname bibnamefont\endcsname\relax
  \def\bibnamefont#1{#1}\fi
\expandafter\ifx\csname bibfnamefont\endcsname\relax
  \def\bibfnamefont#1{#1}\fi
\expandafter\ifx\csname citenamefont\endcsname\relax
  \def\citenamefont#1{#1}\fi
\expandafter\ifx\csname url\endcsname\relax
  \def\url#1{\texttt{#1}}\fi
\expandafter\ifx\csname urlprefix\endcsname\relax\def\urlprefix{URL }\fi
\providecommand{\bibinfo}[2]{#2}
\providecommand{\eprint}[2][]{\url{#2}}

\bibitem[{\citenamefont{Anderson}(1967)}]{anderson1967}
\bibinfo{author}{\bibfnamefont{P.~W.} \bibnamefont{Anderson}},
  \bibinfo{journal}{Phys. Rev. Lett.} \textbf{\bibinfo{volume}{18}},
  \bibinfo{pages}{1049} (\bibinfo{year}{1967}).

\bibitem[{\citenamefont{Tanabe and Ohtaka}(1985)}]{tanabe1985}
\bibinfo{author}{\bibfnamefont{Y.}~\bibnamefont{Tanabe}} \bibnamefont{and}
  \bibinfo{author}{\bibfnamefont{K.}~\bibnamefont{Ohtaka}},
  \bibinfo{journal}{Phys. Rev. B} \textbf{\bibinfo{volume}{32}},
  \bibinfo{pages}{2036} (\bibinfo{year}{1985}).

\bibitem[{\citenamefont{Hentschel and Guinea}(2007)}]{hentschel2007}
\bibinfo{author}{\bibfnamefont{M.}~\bibnamefont{Hentschel}} \bibnamefont{and}
  \bibinfo{author}{\bibfnamefont{F.}~\bibnamefont{Guinea}},
  \bibinfo{journal}{Phys. Rev. B} \textbf{\bibinfo{volume}{76}},
  \bibinfo{pages}{115407} (\bibinfo{year}{2007}).

\bibitem[{\citenamefont{Goold et~al.}(2011)\citenamefont{Goold, Fogarty,
  Lo~Gullo, Paternostro, and Busch}}]{goold2011}
\bibinfo{author}{\bibfnamefont{J.}~\bibnamefont{Goold}},
  \bibinfo{author}{\bibfnamefont{T.}~\bibnamefont{Fogarty}},
  \bibinfo{author}{\bibfnamefont{N.}~\bibnamefont{Lo~Gullo}},
  \bibinfo{author}{\bibfnamefont{M.}~\bibnamefont{Paternostro}},
  \bibnamefont{and} \bibinfo{author}{\bibfnamefont{T.}~\bibnamefont{Busch}},
  \bibinfo{journal}{Phys. Rev. A} \textbf{\bibinfo{volume}{84}},
  \bibinfo{pages}{063632} (\bibinfo{year}{2011}).

\bibitem[{\citenamefont{Knap et~al.}(2012)\citenamefont{Knap, Shashi, Nishida,
  Imambekov, Abanin, and Demler}}]{knap2012}
\bibinfo{author}{\bibfnamefont{M.}~\bibnamefont{Knap}},
  \bibinfo{author}{\bibfnamefont{A.}~\bibnamefont{Shashi}},
  \bibinfo{author}{\bibfnamefont{Y.}~\bibnamefont{Nishida}},
  \bibinfo{author}{\bibfnamefont{A.}~\bibnamefont{Imambekov}},
  \bibinfo{author}{\bibfnamefont{D.~A.} \bibnamefont{Abanin}},
  \bibnamefont{and} \bibinfo{author}{\bibfnamefont{E.}~\bibnamefont{Demler}},
  \bibinfo{journal}{Phys. Rev. X} \textbf{\bibinfo{volume}{2}},
  \bibinfo{pages}{041020} (\bibinfo{year}{2012}).

\bibitem[{\citenamefont{Sindona et~al.}(2013)\citenamefont{Sindona, Goold,
  Lo~Gullo, Lorenzo, and Plastina}}]{sindona2013}
\bibinfo{author}{\bibfnamefont{A.}~\bibnamefont{Sindona}},
  \bibinfo{author}{\bibfnamefont{J.}~\bibnamefont{Goold}},
  \bibinfo{author}{\bibfnamefont{N.}~\bibnamefont{Lo~Gullo}},
  \bibinfo{author}{\bibfnamefont{S.}~\bibnamefont{Lorenzo}}, \bibnamefont{and}
  \bibinfo{author}{\bibfnamefont{F.}~\bibnamefont{Plastina}},
  \bibinfo{journal}{Phys. Rev. Lett.} \textbf{\bibinfo{volume}{111}},
  \bibinfo{pages}{165303} (\bibinfo{year}{2013}).

\bibitem[{\citenamefont{Campbell et~al.}(2014)\citenamefont{Campbell,
  Garc\'{\i}a-March, Fogarty, and Busch}}]{campbell2014}
\bibinfo{author}{\bibfnamefont{S.}~\bibnamefont{Campbell}},
  \bibinfo{author}{\bibfnamefont{M.~A.} \bibnamefont{Garc\'{\i}a-March}},
  \bibinfo{author}{\bibfnamefont{T.}~\bibnamefont{Fogarty}}, \bibnamefont{and}
  \bibinfo{author}{\bibfnamefont{T.}~\bibnamefont{Busch}},
  \bibinfo{journal}{Phys. Rev. A} \textbf{\bibinfo{volume}{90}},
  \bibinfo{pages}{013617} (\bibinfo{year}{2014}).

\bibitem[{\citenamefont{Khemani et~al.}(2015)\citenamefont{Khemani,
  Nandkishore, and Sondhi}}]{khemani2015}
\bibinfo{author}{\bibfnamefont{V.}~\bibnamefont{Khemani}},
  \bibinfo{author}{\bibfnamefont{R.}~\bibnamefont{Nandkishore}},
  \bibnamefont{and} \bibinfo{author}{\bibfnamefont{S.}~\bibnamefont{Sondhi}},
  \bibinfo{journal}{Nature Physics} \textbf{\bibinfo{volume}{11}},
  \bibinfo{pages}{560} (\bibinfo{year}{2015}).

\bibitem[{\citenamefont{Deng et~al.}(2015)\citenamefont{Deng, Pixley, Li, and
  Das~Sarma}}]{deng2015}
\bibinfo{author}{\bibfnamefont{D.-L.} \bibnamefont{Deng}},
  \bibinfo{author}{\bibfnamefont{J.~H.} \bibnamefont{Pixley}},
  \bibinfo{author}{\bibfnamefont{X.}~\bibnamefont{Li}}, \bibnamefont{and}
  \bibinfo{author}{\bibfnamefont{S.}~\bibnamefont{Das~Sarma}},
  \bibinfo{journal}{Phys. Rev. B} \textbf{\bibinfo{volume}{92}},
  \bibinfo{pages}{220201} (\bibinfo{year}{2015}).

\bibitem[{\citenamefont{Cosco et~al.}(2018)\citenamefont{Cosco, Borrelli,
  Laine, Pascazio, Scardicchio, and Maniscalco}}]{coscostoc}
\bibinfo{author}{\bibfnamefont{F.}~\bibnamefont{Cosco}},
  \bibinfo{author}{\bibfnamefont{M.}~\bibnamefont{Borrelli}},
  \bibinfo{author}{\bibfnamefont{E.-M.} \bibnamefont{Laine}},
  \bibinfo{author}{\bibfnamefont{S.}~\bibnamefont{Pascazio}},
  \bibinfo{author}{\bibfnamefont{A.}~\bibnamefont{Scardicchio}},
  \bibnamefont{and}
  \bibinfo{author}{\bibfnamefont{S.}~\bibnamefont{Maniscalco}},
  \bibinfo{journal}{New Journal of Physics} \textbf{\bibinfo{volume}{20}},
  \bibinfo{pages}{073041} (\bibinfo{year}{2018}).

\bibitem[{\citenamefont{Aubry and Andr\'e}(1980)}]{aubry1980}
\bibinfo{author}{\bibfnamefont{S.}~\bibnamefont{Aubry}} \bibnamefont{and}
  \bibinfo{author}{\bibfnamefont{G.}~\bibnamefont{Andr\'e}},
  \bibinfo{journal}{Ann. Israel Phys. Soc} \textbf{\bibinfo{volume}{3}},
  \bibinfo{pages}{18} (\bibinfo{year}{1980}).

\bibitem[{\citenamefont{Sokoloff}(1981)}]{sokoloff1981}
\bibinfo{author}{\bibfnamefont{J.~B.} \bibnamefont{Sokoloff}},
  \bibinfo{journal}{Phys. Rev. B} \textbf{\bibinfo{volume}{23}},
  \bibinfo{pages}{6422} (\bibinfo{year}{1981}).

\bibitem[{\citenamefont{Thouless and Niu}(1983)}]{thouless1983}
\bibinfo{author}{\bibfnamefont{D.~J.} \bibnamefont{Thouless}} \bibnamefont{and}
  \bibinfo{author}{\bibfnamefont{Q.}~\bibnamefont{Niu}},
  \bibinfo{journal}{Journal of Physics A: Mathematical and General}
  \textbf{\bibinfo{volume}{16}}, \bibinfo{pages}{1911} (\bibinfo{year}{1983}).

\bibitem[{\citenamefont{Jitomirskaya}(1999)}]{jitomirskaya1999}
\bibinfo{author}{\bibfnamefont{S.~Y.} \bibnamefont{Jitomirskaya}},
  \bibinfo{journal}{Annals of Mathematics} \textbf{\bibinfo{volume}{150}},
  \bibinfo{pages}{1159} (\bibinfo{year}{1999}).

\bibitem[{\citenamefont{Modugno}(2009)}]{modugno2009}
\bibinfo{author}{\bibfnamefont{M.}~\bibnamefont{Modugno}},
  \bibinfo{journal}{New Journal of Physics} \textbf{\bibinfo{volume}{11}},
  \bibinfo{pages}{033023} (\bibinfo{year}{2009}).

\bibitem[{\citenamefont{Settino et~al.}(2017)\citenamefont{Settino, Lo~Gullo,
  Sindona, Goold, and Plastina}}]{settino2017}
\bibinfo{author}{\bibfnamefont{J.}~\bibnamefont{Settino}},
  \bibinfo{author}{\bibfnamefont{N.}~\bibnamefont{Lo~Gullo}},
  \bibinfo{author}{\bibfnamefont{A.}~\bibnamefont{Sindona}},
  \bibinfo{author}{\bibfnamefont{J.}~\bibnamefont{Goold}}, \bibnamefont{and}
  \bibinfo{author}{\bibfnamefont{F.}~\bibnamefont{Plastina}},
  \bibinfo{journal}{Phys. Rev. A} \textbf{\bibinfo{volume}{95}},
  \bibinfo{pages}{033605} (\bibinfo{year}{2017}).

\bibitem[{\citenamefont{Sanchez-Palencia and
  Lewenstein}(2010)}]{sanchez-palencia2010}
\bibinfo{author}{\bibfnamefont{L.}~\bibnamefont{Sanchez-Palencia}}
  \bibnamefont{and}
  \bibinfo{author}{\bibfnamefont{M.}~\bibnamefont{Lewenstein}},
  \bibinfo{journal}{Nature Physics} \textbf{\bibinfo{volume}{6}},
  \bibinfo{pages}{87 EP } (\bibinfo{year}{2010}).

\bibitem[{\citenamefont{Schreiber et~al.}(2015)\citenamefont{Schreiber,
  Hodgman, Bordia, L{\"u}schen, Fischer, Vosk, Altman, Schneider, and
  Bloch}}]{schreiber2015}
\bibinfo{author}{\bibfnamefont{M.}~\bibnamefont{Schreiber}},
  \bibinfo{author}{\bibfnamefont{S.~S.} \bibnamefont{Hodgman}},
  \bibinfo{author}{\bibfnamefont{P.}~\bibnamefont{Bordia}},
  \bibinfo{author}{\bibfnamefont{H.~P.} \bibnamefont{L{\"u}schen}},
  \bibinfo{author}{\bibfnamefont{M.~H.} \bibnamefont{Fischer}},
  \bibinfo{author}{\bibfnamefont{R.}~\bibnamefont{Vosk}},
  \bibinfo{author}{\bibfnamefont{E.}~\bibnamefont{Altman}},
  \bibinfo{author}{\bibfnamefont{U.}~\bibnamefont{Schneider}},
  \bibnamefont{and} \bibinfo{author}{\bibfnamefont{I.}~\bibnamefont{Bloch}},
  \bibinfo{journal}{Science} \textbf{\bibinfo{volume}{349}},
  \bibinfo{pages}{842} (\bibinfo{year}{2015}).

\bibitem[{\citenamefont{L\"uschen et~al.}(2017)\citenamefont{L\"uschen, Bordia,
  Scherg, Alet, Altman, Schneider, and Bloch}}]{luschen2017}
\bibinfo{author}{\bibfnamefont{H.~P.} \bibnamefont{L\"uschen}},
  \bibinfo{author}{\bibfnamefont{P.}~\bibnamefont{Bordia}},
  \bibinfo{author}{\bibfnamefont{S.}~\bibnamefont{Scherg}},
  \bibinfo{author}{\bibfnamefont{F.}~\bibnamefont{Alet}},
  \bibinfo{author}{\bibfnamefont{E.}~\bibnamefont{Altman}},
  \bibinfo{author}{\bibfnamefont{U.}~\bibnamefont{Schneider}},
  \bibnamefont{and} \bibinfo{author}{\bibfnamefont{I.}~\bibnamefont{Bloch}},
  \bibinfo{journal}{Phys. Rev. Lett.} \textbf{\bibinfo{volume}{119}},
  \bibinfo{pages}{260401} (\bibinfo{year}{2017}).

\bibitem[{\citenamefont{Dorner et~al.}(2012)\citenamefont{Dorner, Goold,
  Cormick, Paternostro, and Vedral}}]{dorner2012}
\bibinfo{author}{\bibfnamefont{R.}~\bibnamefont{Dorner}},
  \bibinfo{author}{\bibfnamefont{J.}~\bibnamefont{Goold}},
  \bibinfo{author}{\bibfnamefont{C.}~\bibnamefont{Cormick}},
  \bibinfo{author}{\bibfnamefont{M.}~\bibnamefont{Paternostro}},
  \bibnamefont{and} \bibinfo{author}{\bibfnamefont{V.}~\bibnamefont{Vedral}},
  \bibinfo{journal}{Phys. Rev. Lett.} \textbf{\bibinfo{volume}{109}},
  \bibinfo{pages}{160601} (\bibinfo{year}{2012}).

\bibitem[{\citenamefont{Mascarenhas et~al.}(2014)\citenamefont{Mascarenhas,
  Bragan\ifmmode~\mbox{\c{c}}\else \c{c}\fi{}a, Dorner, Fran\ifmmode
  \mbox{\c{c}}\else~\c{c}\fi{}a Santos, Vedral, Modi, and
  Goold}}]{mascarenhas2014}
\bibinfo{author}{\bibfnamefont{E.}~\bibnamefont{Mascarenhas}},
  \bibinfo{author}{\bibfnamefont{H.}~\bibnamefont{Bragan\ifmmode~\mbox{\c{c}}\else
  \c{c}\fi{}a}}, \bibinfo{author}{\bibfnamefont{R.}~\bibnamefont{Dorner}},
  \bibinfo{author}{\bibfnamefont{M.}~\bibnamefont{Fran\ifmmode
  \mbox{\c{c}}\else~\c{c}\fi{}a Santos}},
  \bibinfo{author}{\bibfnamefont{V.}~\bibnamefont{Vedral}},
  \bibinfo{author}{\bibfnamefont{K.}~\bibnamefont{Modi}}, \bibnamefont{and}
  \bibinfo{author}{\bibfnamefont{J.}~\bibnamefont{Goold}},
  \bibinfo{journal}{Phys. Rev. E} \textbf{\bibinfo{volume}{89}},
  \bibinfo{pages}{062103} (\bibinfo{year}{2014}).

\bibitem[{\citenamefont{Sharma and Dutta}(2015)}]{sharma2015}
\bibinfo{author}{\bibfnamefont{S.}~\bibnamefont{Sharma}} \bibnamefont{and}
  \bibinfo{author}{\bibfnamefont{A.}~\bibnamefont{Dutta}},
  \bibinfo{journal}{Phys. Rev. E} \textbf{\bibinfo{volume}{92}},
  \bibinfo{pages}{022108} (\bibinfo{year}{2015}).

\bibitem[{\citenamefont{Bayocboc and Paraan}(2015)}]{bayocboc2015}
\bibinfo{author}{\bibfnamefont{F.~A.} \bibnamefont{Bayocboc}} \bibnamefont{and}
  \bibinfo{author}{\bibfnamefont{F.~N.~C.} \bibnamefont{Paraan}},
  \bibinfo{journal}{Phys. Rev. E} \textbf{\bibinfo{volume}{92}},
  \bibinfo{pages}{032142} (\bibinfo{year}{2015}).

\bibitem[{\citenamefont{Bayat et~al.}(2016)\citenamefont{Bayat, Apollaro,
  Paganelli, De~Chiara, Johannesson, Bose, and Sodano}}]{bayat2016}
\bibinfo{author}{\bibfnamefont{A.}~\bibnamefont{Bayat}},
  \bibinfo{author}{\bibfnamefont{T.~J.~G.} \bibnamefont{Apollaro}},
  \bibinfo{author}{\bibfnamefont{S.}~\bibnamefont{Paganelli}},
  \bibinfo{author}{\bibfnamefont{G.}~\bibnamefont{De~Chiara}},
  \bibinfo{author}{\bibfnamefont{H.}~\bibnamefont{Johannesson}},
  \bibinfo{author}{\bibfnamefont{S.}~\bibnamefont{Bose}}, \bibnamefont{and}
  \bibinfo{author}{\bibfnamefont{P.}~\bibnamefont{Sodano}},
  \bibinfo{journal}{Phys. Rev. B} \textbf{\bibinfo{volume}{93}},
  \bibinfo{pages}{201106} (\bibinfo{year}{2016}).

\bibitem[{\citenamefont{Paganelli and Apollaro}(2016)}]{paganelli2016}
\bibinfo{author}{\bibfnamefont{S.}~\bibnamefont{Paganelli}} \bibnamefont{and}
  \bibinfo{author}{\bibfnamefont{T.~J.} \bibnamefont{Apollaro}},
  \bibinfo{journal}{International Journal of Modern Physics B} p.
  \bibinfo{pages}{1750065} (\bibinfo{year}{2016}).

\bibitem[{\citenamefont{Cosco et~al.}(2017)\citenamefont{Cosco, Borrelli,
  Silvi, Maniscalco, and De~Chiara}}]{coscocoulomb}
\bibinfo{author}{\bibfnamefont{F.}~\bibnamefont{Cosco}},
  \bibinfo{author}{\bibfnamefont{M.}~\bibnamefont{Borrelli}},
  \bibinfo{author}{\bibfnamefont{P.}~\bibnamefont{Silvi}},
  \bibinfo{author}{\bibfnamefont{S.}~\bibnamefont{Maniscalco}},
  \bibnamefont{and}
  \bibinfo{author}{\bibfnamefont{G.}~\bibnamefont{De~Chiara}},
  \bibinfo{journal}{Phys. Rev. A} \textbf{\bibinfo{volume}{95}},
  \bibinfo{pages}{063615} (\bibinfo{year}{2017}).

\bibitem[{\citenamefont{Rossini and Vicari}(2018)}]{rossini2018}
\bibinfo{author}{\bibfnamefont{D.}~\bibnamefont{Rossini}} \bibnamefont{and}
  \bibinfo{author}{\bibfnamefont{E.}~\bibnamefont{Vicari}},
  \bibinfo{journal}{arXiv preprint arXiv:1807.01674}  (\bibinfo{year}{2018}).

\bibitem[{\citenamefont{Sindona et~al.}(2014)\citenamefont{Sindona, Goold,
  Gullo, and Plastina}}]{sindona2014}
\bibinfo{author}{\bibfnamefont{A.}~\bibnamefont{Sindona}},
  \bibinfo{author}{\bibfnamefont{J.}~\bibnamefont{Goold}},
  \bibinfo{author}{\bibfnamefont{N.~L.} \bibnamefont{Gullo}}, \bibnamefont{and}
  \bibinfo{author}{\bibfnamefont{F.}~\bibnamefont{Plastina}},
  \bibinfo{journal}{New Journal of Physics} \textbf{\bibinfo{volume}{16}},
  \bibinfo{pages}{045013} (\bibinfo{year}{2014}).

\bibitem[{\citenamefont{Talkner et~al.}(2007)\citenamefont{Talkner, Lutz, and
  H\"anggi}}]{talkner2007}
\bibinfo{author}{\bibfnamefont{P.}~\bibnamefont{Talkner}},
  \bibinfo{author}{\bibfnamefont{E.}~\bibnamefont{Lutz}}, \bibnamefont{and}
  \bibinfo{author}{\bibfnamefont{P.}~\bibnamefont{H\"anggi}},
  \bibinfo{journal}{Phys. Rev. E} \textbf{\bibinfo{volume}{75}},
  \bibinfo{pages}{050102} (\bibinfo{year}{2007}).

\bibitem[{\citenamefont{Plastina et~al.}(2014)\citenamefont{Plastina, Alecce,
  Apollaro, Falcone, Francica, Galve, Lo~Gullo, and Zambrini}}]{plastina2014}
\bibinfo{author}{\bibfnamefont{F.}~\bibnamefont{Plastina}},
  \bibinfo{author}{\bibfnamefont{A.}~\bibnamefont{Alecce}},
  \bibinfo{author}{\bibfnamefont{T.~J.~G.} \bibnamefont{Apollaro}},
  \bibinfo{author}{\bibfnamefont{G.}~\bibnamefont{Falcone}},
  \bibinfo{author}{\bibfnamefont{G.}~\bibnamefont{Francica}},
  \bibinfo{author}{\bibfnamefont{F.}~\bibnamefont{Galve}},
  \bibinfo{author}{\bibfnamefont{N.}~\bibnamefont{Lo~Gullo}}, \bibnamefont{and}
  \bibinfo{author}{\bibfnamefont{R.}~\bibnamefont{Zambrini}},
  \bibinfo{journal}{Phys. Rev. Lett.} \textbf{\bibinfo{volume}{113}},
  \bibinfo{pages}{260601} (\bibinfo{year}{2014}).

\bibitem[{\citenamefont{Campisi and Fazio}(2016)}]{campisi2016}
\bibinfo{author}{\bibfnamefont{M.}~\bibnamefont{Campisi}} \bibnamefont{and}
  \bibinfo{author}{\bibfnamefont{R.}~\bibnamefont{Fazio}},
  \bibinfo{journal}{Nature Communications} \textbf{\bibinfo{volume}{7}},
  \bibinfo{pages}{11895 EP } (\bibinfo{year}{2016}).

\end{thebibliography}

\end{document}